# Icosahedral quasicrystal, 1/1 and 2/1 approximants in Zn-based ternary alloys containing Au and Yb/Tb

Tsutomu Ishimasa*

*Toyota Physical & Chemical Research Institute, Nagakute, Aichi, 480-1192 Japan*

**Abstract**

In a narrow composition range centered at $Zn_{74.5}Au_{10.5}Yb_{15.0}$, Tsai-type icosahedral quasicrystal is formed in alloys quenched from 880 ºC. This quasicrystal belongs to the primitive type with the 6-dimensional lattice parameter $a_{6D}$=7.378 Å. The quasicrystal was not formed in the slowly cooled specimen, and was considered a metastable phase. The stable phase is a 2/1 approximant of the lattice parameter $a_{2/1}$=23.271 Å. This approximant forms exclusively in $Zn_{76.0}Au_{9.0}Yb_{15.0}$ alloy annealed at 530 ºC. In addition, $Zn_{70.5}Au_{15.5}Tb_{14.0}$ alloy annealed at 505 ºC forms a Tsai-type 1/1 approximant ($a_{1/1}$=14.343 Å). These new Zn-based phases observed in this study correspond to the quasicrystal-related phases in binary Cd-lanthanoid systems, and show the possibility of isostructural substitution of Cd by Zn/Au.

*Corresponding Author

   E-mail :   ishimasa@toyotariken.jp

   Telephone:   +81-561-57-9523

   Fax:        +81-561-63-6302

1. **Introduction**

Intensive research over the past 20 years has revealed that there are more than 40 new icosahedral quasicrystals in binary and ternary alloy systems [1, 2]. Almost all fall into one structure type named Tsai-type. Here the classification is based on the details of the local structural unit, *i.e.* cluster. In the case of Tsai-type, isostructural substitution has been effectively applied, and the $Cd_{5.7}Yb$ quasicrystal [3] is a starting point. All known Tsai-type quasicrystals can be considered as its derivatives.

At present, formation conditions for Tsai-type quasicrystals are known empirically. One is electronic condition, and the other is atomic size condition. The former is related to a specific average number of valence electron par atom, *e/a*, and is related to the stabilization mechanism of the electronic system by the interaction between the Fermi surface and the pseudo-Brillouin zone [4-6]. In the case of Tsai-type quasicrystals, it is known that the value *e/a* is in the range of 2.0 to 2.1. The latter condition concerns the atomic size ratio $\eta$ between the larger and smaller elements. For example, for $Cd_{5.7}Yb$ $\eta = r_{Yb}/r_{Cd}$, where $r_{Yb}$ and $r_{Cd}$ are the atomic radii of Yb and Cd, respectively. This ratio is related to the sphere-packing conditions for the construction of the triple-shell of Tsai-type cluster [7]. The ideal ratio calculated geometrically is $\eta = 1.288$. These two conditions with respect to *e/a* and $\eta$ correspond to the Hume-Rothery rules modified for quasicrystals. Note that approximants, which are periodic analogs of quasicrystals composed of the same type of cluster, satisfy very similar formation conditions.

By using such formation conditions as substitution rules, it has become possible to synthesize new materials with interesting physical properties. The Au-Al-Yb magnetic quasicrystal that exhibits quantum criticality is an example [8]. Yb in this quasicrystal has an intermediate valence. In addition to this, several icosahedral quasicrystals containing Yb are known. They are Zn-Mg-Yb [7], Ag-In-Yb [9], $Cd_{5.7}Yb$ [3], Cd-Mg-Yb [10] and Au-Sn-Yb [11]. However, at ambient pressure Yb is divalent or nearly divalent in all cases. Another interesting phenomenon is antiferromagnetic ordering of Tsai-type 1/1 approximants, $Cd_6Tb$ [12] and Au-Al-*L* (*L*=Gd and Tb) [13]. Lanthanoid atoms with localized magnetic moments are located at the twelve vertices of the icosahedral shell in these Tsai-type approximants. Although the arrangement of magnetic atoms is similar in the corresponding quasicrystals, no such magnetic order has yet been realized in them.

To elucidate the physical properties intrinsic to quasicrystals, it is important to find new alloy systems that contain lanthanoids in particular. The purpose of this study is to expand the alloy systems in which quasicrystal-related phases are formed. In this letter, we will report the experimental results of the following two ternary systems: Zn-Au-Yb and Zn-Au-Tb with approximately 15 at.% lanthanoids.

## 2. Experimental procedures

Specimens of Zn-based alloys were prepared using high-purity materials of Zn (Nilaco; 99.99 %), Au (Tanaka; 99.99 %), Tb (Rare metallic; 99.9 %) and Yb (Rare metallic; 99.9 %). The material weighted in the appropriate composition was placed in a pure alumina crucible, and sealed in a silica ampule after evacuating to a pressure of approximately $1.0 \times 10^{-6}$ Torr. The total weight of the alloy specimen ranges from 3.2 to 7.4 g. A piece of Ti wire (diameter 1mm, typical weight 100 mg) was put in the silica ampule as a getter. For the heat treatment, an electric furnace with a controlled temperature gradient was used to suppress the evaporation of Zn. See [6] for more details. The Zn-Au-Yb specimen was melted in two stages: 590 ºC for 5 hours and 850-880 ºC for 15 hours. After the melting process, two types of heat treatment were applied. One is quenching and the other is slow cooling. In the former case, the silica ampule was cooled with water from the melting temperature, but the ampule was not broken. In the latter case, the specimen was cooled to 530 ºC at a rate of -16 K/h, and annealed at this temperature for tens of hours. The Zn-Au-Tb specimen was further melted at 950 ºC after the two-step melting, and cooled to 505 ºC at a rate of -22 K/h. The specimen was annealed at this temperature for 30 h.

Phase characterization was performed by powder X-ray diffraction using Cu K$\alpha$ radiation (Rigaku RINT-2100), and selected-area electron diffraction (JEOL JEM-200CS: acceleration voltage 160kV). After observing the morphology with a scanning electron microscope (Thermo Fisher Scientific PRISMA E), the composition of each phase was analyzed by EDS (energy dispersive spectroscopy). RIETAN-2000 developed by Izumi and Ikeda [14] was used to simulate X-ray diffraction pattern and also to perform Rietveld refinement. The software, Balls & Sticks [15], was used for graphical visualization of structure model.

## 3. Results and Discussion

In the quenched alloys with the composition $Zn_{100-x-y}Au_xYb_y$, a P-type icosahedral quasicrystal is formed, where 9≤x≤11 and 14≤y≤16. In the $Zn_{74.5}Au_{10.5}Yb_{15.0}$ alloy, the quasicrystal was formed exclusively. The powder X-ray diffraction pattern of this alloy is shown in Figure 1a. Almost all reflections are indexed using six integers according to the scheme proposed by Elser [16]. No reflection condition was detected in the indices, which indicates that this icosahedral quasicrystal is of primitive type, namely P-type. The 6-dimensional lattice parameter $a_{6D}$ was estimated to be 7.378(2) Å. Typical electron diffraction patterns are presented in Figures 2a~c, which show the icosahedral symmetry, $m\bar{3}\bar{5}$. Overall, the reflection arrangement satisfies the scaling law at the third power of the golden ratio. For example, one notices it in the direction of the 5-fold axis in Figure 2a. This scaling property again indicates the presence of P-type quasicrystal. The quasicrystal composition was analyzed as $Zn_{72.8(4)}Au_{10.6(2)}Yb_{16.6(3)}$ by EDS. This is in agreement with the nominal one. The quasicrystal survived after aging at 600 °C for 108 hours, but did not form in alloys that were slowly cooled from the melt. Therefore, the quasicrystal is considered a metastable phase. In the quasicrystal specimens, the following three phases were observed as minor phases: ZnAuYb-type (space group: *Pnma*), $Zn_{58}Yb_{13}$-type (space group: $P6_3/mmc$) and the 2/1 cubic approximant (space group: $Pa\bar{3}$) described below.

In the $Zn_{76.0}Au_{9.0}Yb_{15.0}$ alloy annealed at 530 °C, the 2/1 approximant was formed exclusively, of which powder X-ray diffraction pattern is presented in Figure 1b. The lattice parameter $a_{2/1}$ is 23.271(2) Å that can be related to the 6-dimmensional lattice parameter $a_{6D}$= 7.378 Å of the icosahedral quasicrystal by the following equation [17] with consecutive Fibonacci numbers, where $F_2$=2 and $F_1$=1.

$$a_{n+1/n} = \sqrt{\frac{2}{2+\tau}}\left(F_n + \tau F_{n+1}\right)a_{6D},$$

where $a_{n+1/n}$ is the lattice parameter of $n+1/n$ approximant, and $\tau$ is the golden ratio (1+√5)/2.

Figures 2d~f show the electron diffraction patters of the 2/1 approximant. The intensity distributions of the patterns are similar to those of the quasicrystal as seen in Figures 2a and d, as well as b and e. The intensities of the following reflections are relatively strong; 850 reflection in Figure 2d, and 352 and 583 in Figure 2e. These reflections have a Miller index

consisting of nearby Fibonacci numbers. This is one reason for the similarity in intensity distribution between the quasicrystal and approximant. The space group of the approximant was determined to be $Pa\bar{3}$, because there are three 2-fold axes orthogonal to each other, 3-fold axes and the reflection condition $0kl$ for even $k$. By EDS, the composition of the approximant was analyzed as $Zn_{76.5(2)}Au_{8.5(1)}Yb_{15.0(2)}$.

In order to demonstrate that this approximant is composed of Tsai-type cluster, the powder X-ray diffraction pattern shown in Figure 1c was calculated assuming the following structure model. (Note that the 2/1 approximant contains about 700 atoms in the unit cell and is too complex to perform Rietveld refinement using normal diffraction data.) This is based on the structure model of Zn-Mg-Sc 2/1 approximant proposed by Lin and Corbett [18]. In this calculation, the average atom $M=0.894Zn+0.106Au$ was used instead of individual Zn and Au. In addition, Zn/Mg and Sc in their model were replaced with the mean atom $M$ and Yb, respectively. The resulting pattern is presented in Figure 1c. The overall agreement between Figures 1b and c certainly shows that the 2/1 approximant is classified as Tsai-type. Therefore, it is natural to assume that the icosahedral quasicrystal is also Tsai-type.

In the case of Zn-Au-Tb system, a 1/1 approximant is formed. Figure 3 shows powder X-ray diffraction pattern of $Zn_{70.5}Au_{15.5}Tb_{14.0}$ alloy annealed at 505 ºC for 30 h. The unit cell is a body-centered cubic with a lattice parameter $a_{1/1}=14.343$ (2) Å. Rietveld refinement smoothly converged by assuming the structure model (space group: $Im\bar{3}$) with Tsai-type cluster. The reliable factors were $R_{WP}=8.62\%$, $R_I=1.97\%$ and $R_F=1.18\%$.

The result of the refinement is shown in Table I and Figure 4. The unit cell contains 122.3 Zn, 29.7 Au and 24.0 Tb atoms, for a total of 176.0 atoms. The composition of this model is $Zn_{69.5}Au_{16.9}Tb_{13.6}$. This is consistent with the nominal composition $Zn_{70.5}Au_{15.5}Tb_{14.0}$ as well as $Zn_{68.9(4)}Au_{16.3(1)}Tb_{14.8(3)}$ analyzed by EDS. There are nine crystallographic sites: M1-M8 and Tb sites. The Au atoms are mainly located at the following three sites: M2, M3 and M7. The following four sites, M4, M5, M6 and M8 are solely occupied by Zn, and the site M6 slightly splits into two as presented in Figure 4c.

This crystal structure is a typical example of the 1/1 approximant of Tsai-type except for the split of M6. Tsai-type clusters shown in Figure 4c are embedded in body-centered network formed by M3 and M5 sites (Figure 4d). The Tsai-type cluster consists of a triple shell with an irregular central region (Figure 4a) in which four atoms form a randomly orientated tetrahedron [19]. The triple shell consists of the innermost dodecahedron (M2 and M4), icosahedron (Tb), and icosidodecahedron (M1, M6). The Zn atom at the M8 site connects two Tsai-type clusters along the [111] direction. This result confirms that the Tsai-type 1/1 approximant is formed in Zn-Au-Tb system. It is noted that no corresponding approximant has been reported in binary Zn-Tb system.

For binary Zn alloys, two approximants of 1/1 $Zn_6Sc$ and $Zn_6Yb$ ($Zn_{17}Yb_3$) are known [20]. Both are composed of Tsai-type clusters, but the central structure of the clusters is different. In the case of $Zn_6Yb$, the central region is almost occupied by a single Yb atom, but in $Zn_6Sc$, Zn tetrahedrons with a random orientation are arranged like in the Zn-Au-Tb approximant. The size relationship between the rare-earth and Zn atoms affects not only the sphere-packing in the triple shell but also the central structure of the cluster. The atomic size ratio $\eta$ is calculated to be 1.178 and 1.392 for $Zn_6Sc$ and $Zn_6Yb$, respectively. For this calculation, we used the atomic radii $r_{Zn}$=1.394 Å, $r_{Sc}$=1.641 Å, $r_{Yb}$=1.940 Å [22], assuming divalent Yb. The value of $Zn_6Sc$ is smaller than the ideal value of 1.288, but $Zn_6Yb$ is larger. It is interesting to know how this difference affects the ability of Zn-based alloys to form Tsai-type quasicrystals. In relation to the $Zn_6Sc$ approximant, binary Zn-Sc quasicrystal [23] and ternary Zn-$M$-Sc quasicrystals are formed, where $M$=Mg, Fe, Co, Ni, Cu, Ag and Pd [2, 6]. They are all stable phases. However, there is no known stable quasicrystal related to $Zn_6Yb$, but only two metastable quasicrystals. They are $Zn_{76.2}Mg_{9.3}Yb_{14.5}$ [7] with $\eta$=1.369 and $Zn_{74.5}Au_{10.5}Yb_{15.0}$ with $\eta$=1.386, the latter being reported in this letter. Here, the ratio of the average radius of Zn and Mg (Zn and Au) and the radius of Yb was used. The absence of a stable quasicrystal suggests that increasing the atomic size ratio $\eta$ is disadvantageous for forming a stable phase. (The size ratios $\eta$ of the $Zn_{76.0}Au_{9.0}Yb_{15.0}$ 2/1 approximant and $Zn_{70.5}Au_{15.5}Tb_{14.0}$ 1/1 approximant are 1.387 and 1.270, respectively.) On the other hand, in terms of the intermediate valence of Yb, the combination of atoms with a large $\eta$ has the advantage of generating the intermediate valence with the help of chemical pressure [21]. This is because the atomic size of trivalent Yb is smaller than that of divalent Yb. In this meaning, measuring of the valence of Yb in the new Zn-Au-Yb quasicrystal is an interesting future subject.

To clarify the details of the central structure, it is important to analyze the structure of the Zn-Au-Yb quasicrystal. However, this is difficult to do because of the metastability. A possible way is to treat the stable 2/1 approximant reported in this letter. The structure model used to simulate powder X-ray diffraction (Figure 1c) is based on Zn-Mg-Sc Lin-Corbett model [18]. In this model each cluster contains a Zn/Au tetrahedron. The agreement between the measured and calculated intensities suggests the presence of a tetrahedron in the central region, but future studies are needed to confirm this.

Regarding the valence electron concentration, the $e/a$ values of the three new phases are 1.90 for $Zn_{74.5}Au_{10.5}Yb_{15.0}$ quasicrystal, 1.91 for $Zn_{76.0}Au_{9.0}Yb_{15.0}$ 2/1 approximant, and 1.99 for $Zn_{70.5}Au_{15.5}Tb_{14.0}$ 1/1 approximant. Here the following valences are assumed: 1 for Au, 2 for Zn and Yb, and 3 for Tb. These $e/a$ values are slightly smaller than other Tsai-type quasicrystal-related phases. This is mainly because the valence of Au is smaller than that of Zn. The presence of a phase with a small value of $e/a$ and a large atomic size ratio $\eta$ expands the formation conditions of a Tsai-type quasicrystal-related phase. This is a useful hint to further explore new materials according to the concept of isostructural substitution.

Similar to the Zn-based systems, quasicrystal-related phases are also known in Cd-$L$ ($L$: lanthanoid) binary systems. For example, the Cd-Yb system forms 1/1 and 2/1 approximants with an icosahedral quasicrystal [1]. Their structures are systematically understood as quasiperiodic or periodic arrangements of Tsai-type clusters [24]. In addition, similar phases are generally formed in other Cd-$L$ alloys. As mentioned above, there is no phase corresponding to the Zn-Au-Tb approximant in the Zn-Tb binary system, but in Cd-Tb system. That is $Cd_6Tb$ 1/1 approximant [12, 19]. All three quasicrystal-related phases observed in the present study have corresponding phases in the Cd-$L$ systems. Therefore, it is important to test the possibility of isostructural substitution of Cd by Zn/Au in other Cd-$L$ phases.

4. Conclusion

In this study, a metastable icosahedral quasicrystal and a stable 2/1 approximant were discovered in Zn-Au-Yb system. In addition, a stable 1/1 approximant was observed in Zn-Au-Tb system. These results suggest the existence of undiscovered quasicrystal-related phases in Zn-based ternary alloys containing other lanthanoids. Further research is expected to reveal the unique properties of quasicrystals and quasicrystal-related phases.

**Acknowledgement**

The author thanks N.K. Sato. K. Deguchi and K. Imura for valuable discussions through this study. This work was supported by Kakenhi Grant-in-Aid Nos. 17K05524 and 19H05818 from the Japan Society for the Promotion of Science (JSPS).

**References**

[1] A.P. Tsai, *Discovery of stable icosahedral quasicrystals: progress in understanding structure and properties*, Chem. Soc. Rev., 42 (2013), pp. 5352-5365.

[2] T. Ishimasa, *New group of icosahedral quasicrystals*, Chap. 3 in Quasicrystals, Handbook of metal physics, Elsevier, Amsterdam, 2008, pp. 49-74.

[3] A.P. Tsai, J.Q. Guo, E. Abe, H. Takakura, and T.J. Sato, *A stable binary quasicrystal*, Nature 408 (2000), pp. 537.

[4] A.P. Tsai, *A test of Hume-Rothery rules for stable quasicrystals*, J. Non-crystal. solid, 334-335 (2004), pp. 317-322.

[5] U. Mizutani, *Hume-Rothery rule in structurally complex alloy phases*, The science of complex alloy phases, edited by T.B. Massalski and P.E.A. Turchi, TMS (The minerals, metals and materials Society), 2005, pp. 1-42.

[6] T. Ishimasa, *Hume-Rothery rule as a formation condition of new icosahedral quasicrystals*, *ibid.*, pp. 231-249.

[7] T. Mitani and T. Ishimasa, *A metastable icosahedral quasicrystal in the Zn-Mg-Yb alloy system*, Phil. Mag., 86 (2006), pp. 361-366.

[8] K. Deguchi, N.K. Sato, T. Hattori, K. Ishida, H. Takakura, and T. Ishimasa, *Quantum critical state in a magnetic quasicrystal*, Nat. Mater. 11 (2012), pp. 1013-1016.

[9] J.Q. Guo and A.P. Tsai, *Stable icosahedral quasicrystals in the Ag-In-Ca, Ag-InYb, Ag-In-Ca-Mg, and Ag-In-Yb-Mg systems*, Phil. Mag. Lett. 82 (2002), pp. 349-352.

[10] J.Q. Guo, E. Abe, and A.P. Tsai, *Stable Cd-Mg-Yb and Cd-Mg-Ca icosahedral quasicrystals with wide composition ranges*, Phil. Mag. Lett. 82 (2002), pp. 27-35.

[11] T. Yamada, Y. Nakamura, T. Watanuki, A. Machida, M. Mizukami, K. Nitta, A. Sato, Y. Matsushita, and A.P. Tsai, *Formation of an intermediate valence icosahedral quasicrystal in the Au-Sn-Yb system*, Inorg. Chem. 58 (2019), pp. 9181-9186.

[12] R. Tamura, Y. Muro, T. Hiroto, K. Nishimoto, and T. Takabatake, *Long-range magnetic order in the quasicrystalline approximant $Cd_6Tb$*, Phys. Rev. B 82 (2010), p. 220201(R).

[13] A. Ishikawa, T. Fujii, T. Takeuchi, T. Yamada, Y. Matsushita, and R. Tamura, *Antiferromagnetic order is possible in ternary quasicrystal approximants*, Phys Rev. B 98 (2018), p. 220403(R).

[14] F. Izumi and T. Ikeda, *A. Rietveld-analysis program RIETAN-98 and its applications to zeolites*, Mater. Sci. Forum, 321-324 (2000), pp.198-205.

[15] T.C. Ozawa and S.J. Kang, *Balls & sticks: Easy-to-use structure visualization and*

*animation creating program*, J. Appl. Cryst. 37 (2004), pp. 679.

[16] V. Elser, *The diffraction pattern of projected structures*, Acta Crystallogr. A, 42 (1986), pp. 36-43.

[17] V. Elser and C.L. Henley, *Crystal and quasicrystal structures in Al-Mn-Si alloys*, Phys. Rev. Lett., 55 (1985), pp. 2883-2886.

[18] Q. Lin and J.D. Corbett, *The 1/1 and 2/1 approximants in the Sc-Mg-Zn quasicrystal system: Triacontahedral clusters as fundamental building blocks*, J. Am. Chem. Soc., **128** (2006), pp. 13268-13273.

[19] C.P. Gómez and S. Lidin, *Comparative structural study of the disordered $MCd_6$ quasicrystal approximants*, Phys. Rev. B 68 (2003), p. 24203.

[20] G. Bruzzone, M.L. Fornasini, and F. Merlo, *Rare-earth intermediate phases with zinc*, J. Less-Common Metals 22 (1970), pp. 253-264.

[21] T. Ishimasa, Y. Tanaka, and S. Kashimoto, *Icosahedral quasicrystal and 1/1 cubic approximant in Au-Al-Yb alloys*, Phil. Mag. 91 (2011), pp. 4218-4229.

[22] W.B. Pearson, The Crystal Chemistry and Physics of Metals and Alloys, Wiley, New York, 1972, pp. 151.

[23] P.C. Canfield, M.L. Caudle, C.-S. Ho, A. Kreyssig, S. Nandi, M.G. Kim, X. Lin, A. Kracher, K.W. Dennis, R.W. McCallum, and A.I. Goldman, *Solution growth of a binary icosahedral quasicrystal of $Sc_{12}Zn_{58}$*, Phys. Rev. B 81 (2010), p. 20201(R).

[24] H. Takakura, C.P. Gómez, A. Yamamoto, M. de Boissieu, and A.P. Tsai, *Atomic structure of the binary icosahedral Yb-Cd quasicrystal*, Nature Mater. 6 (2007), pp. 58-63.

**Figure Captions**

Figure 1

Powder X-ray diffraction patterns of Zn-Au-Yb alloys measured with Cu-K$\alpha$ radiation. (a) P-type icosahedral quasicrystal formed in $Zn_{74.5}Au_{10.5}Yb_{15.0}$ alloy quenched from 884 ºC. Reflections due to the icosahedral quasicrystal are indexed using six integers. (b) 2/1 cubic approximant formed in $Zn_{76.0}Au_{9.0}Yb_{15.0}$ alloy annealed at 530 ºC for 67 h. (c) Simulated diffraction pattern of Tsai-type 2/1 cubic approximant.

Figure 2

Electron diffraction patterns of P-type icosahedral quasicrystal, (a)~(c) and 2/1 approximant, (d)~(f) formed in Zn-Au-Yb alloys. (a) 2-fold, (b) 3-fold and (c) 5-fold. (d) [000], (e) [1$\bar{1}$1] and (c) [1$\bar{1}$0]. Arrow in (a) indicates direction of 5-fold axis. Indices of four reflections A~D in (d) and (e) are A: 10 00, B: 850, C: 352, D: 583, respectively.

Figure 3

Powder X-ray diffraction pattern of 1/1 approximant formed in $Zn_{70.5}Au_{15.5}Tb_{14.0}$ alloy. Calculated intensity, measured intensity, peak positions, and deviations are presented in this order.

Figure 4

Structure model of cubic 1/1 approximant of Zn-Au-Tb derived by Rietveld refinement. (a) Central structure of cluster. (b) Dodecahedron with central part. (c) Tsai-type cluster and M8 site. Notice splitting at M6 site. (d) Network formed by M3 and M5 sites.

Table 1. Result of Rietveld refinement of Zn-Au-Tb 1/1 cubic approximant. Occupancy of the site M7 is 1/6.

| Site | Atom | Set | $x$ | $y$ | $z$ | $B$ (Å$^2$) |
|---|---|---|---|---|---|---|
| M1 | 0.903Zn + 0.097Au | 48$h$ | 0.3452(2) | 0.1969(2) | 0.1071(2) | 1.61(8) |
| M2 | 0.592Zn + 0.408Au | 24$g$ | 0 | 0.2460(2) | 0.0884(2) | 0.60(6) |
| M3 | 0.477Zn + 0.523Au | 24$g$ | 0 | 0.5964(2) | 0.6519(1) | 0.43(5) |
| M4 | Zn | 16$f$ | 0.1566(2) | --- | --- | 1.3(1) |
| M5 | Zn | 12$e$ | 0.1956(3) | 0 | 1/2 | 0.51(9) |
| M6 | 0.5Zn | 24$g$ | 0 | 0.4135(6) | 0.0191(8) | 0.9(2) |
| M7 | 0.111Zn + 0.056Au | 48$h$ | 0.026(1) | 0.088(1) | 0.091(1) | 9.3(6) |
| M8 | Zn | 8$c$ | 1/4 | 1/4 | 1/4 | 4.3(2) |
| Tb | Tb | 24$g$ | 0 | 0.1884(1) | 0.3012(1) | 0.62 (4) |

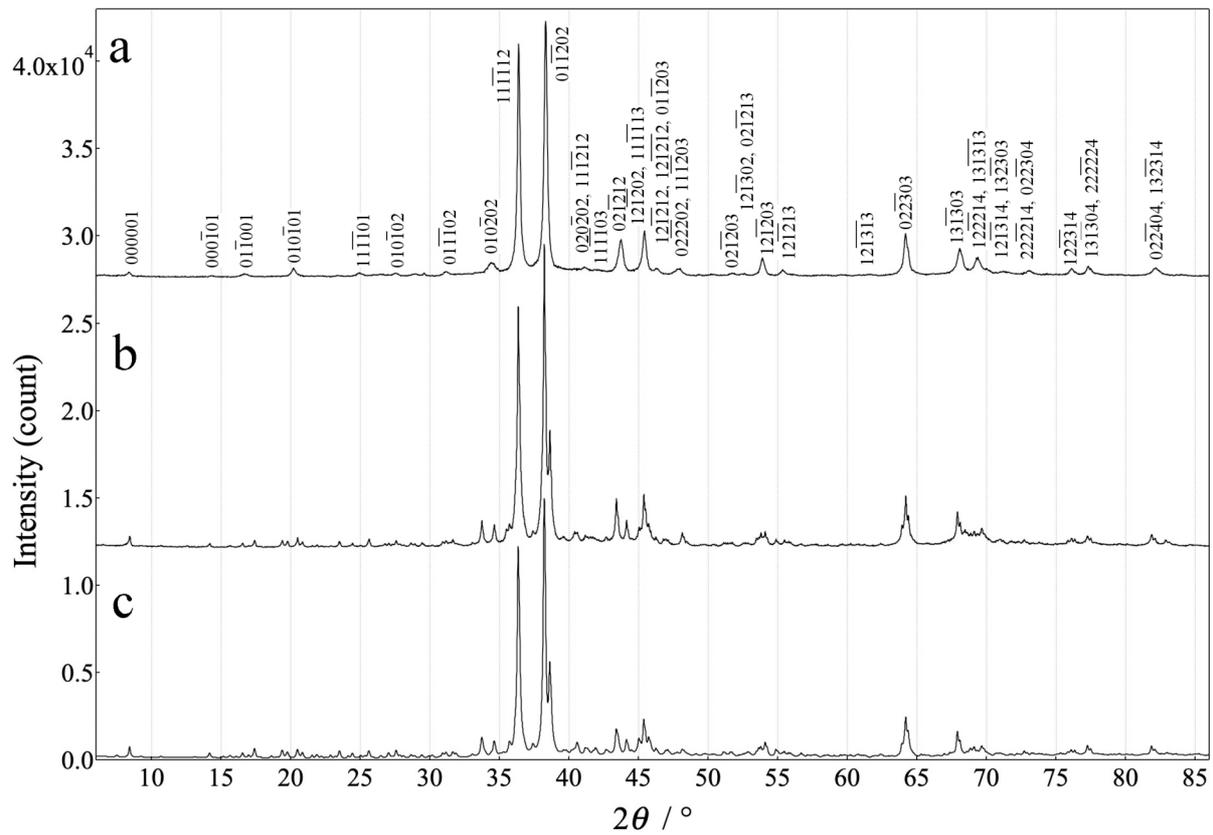

Fig. 1

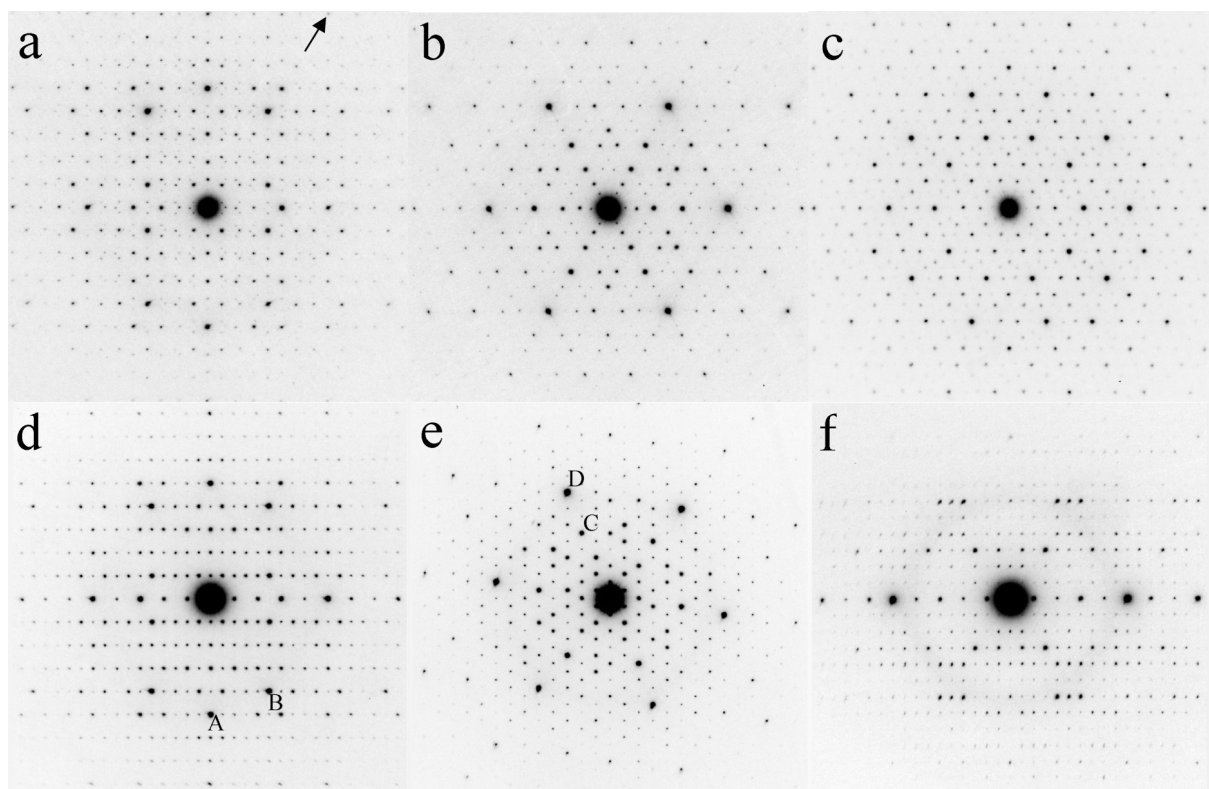

Fig. 2

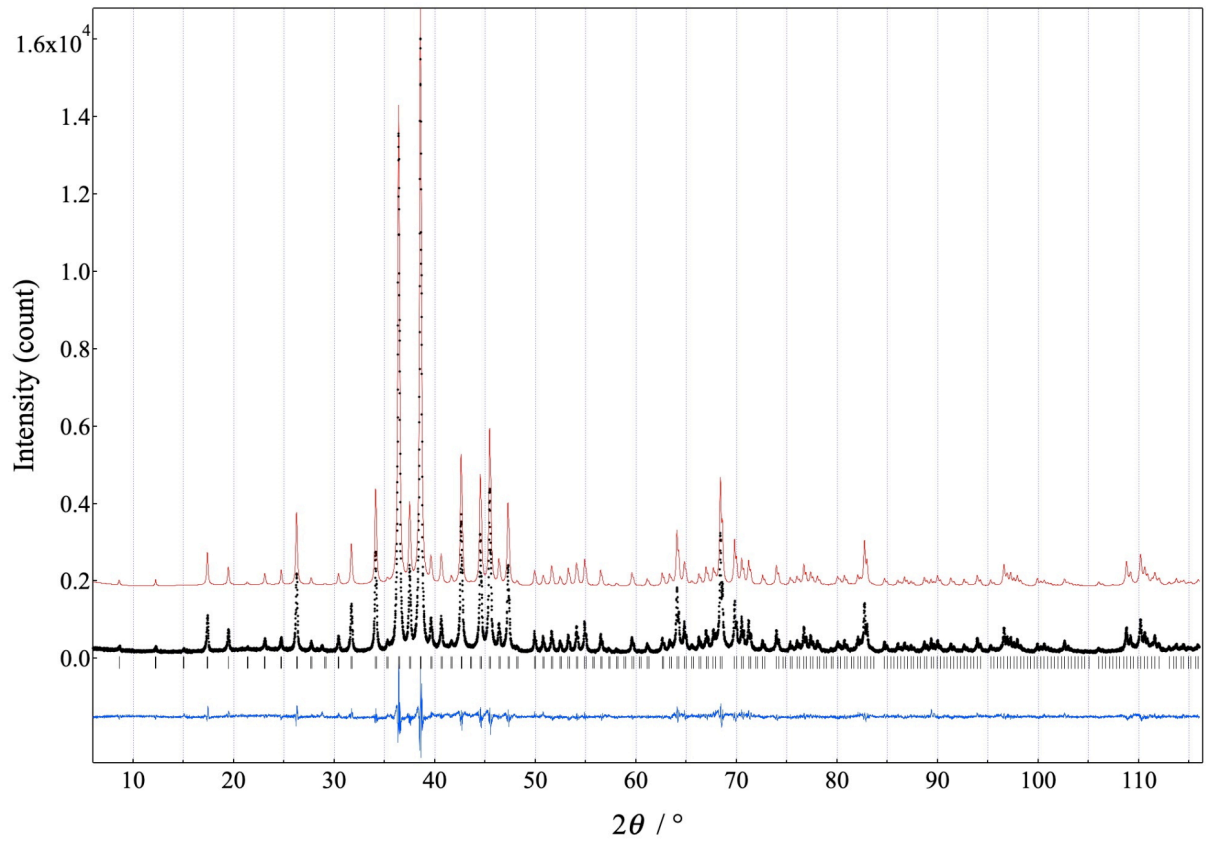

Fig. 3

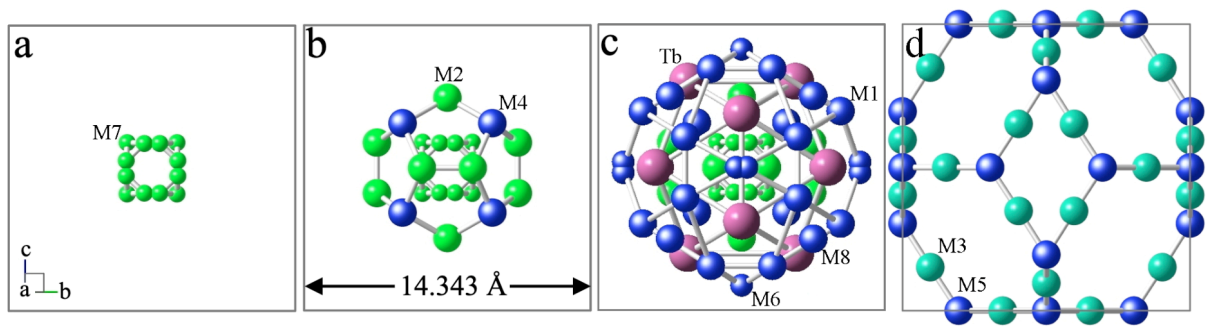

Fig. 4